\documentclass[doublecol]{epl2}

\usepackage{amsfonts}
\usepackage{amsmath}
\usepackage{amssymb}
\usepackage{hyperref}

\hypersetup{
  colorlinks=true,linkcolor=blue,citecolor=blue,
  filecolor=blue,urlcolor=blue,breaklinks=true
}

\title{Remarks on the Aharonov-Casher dynamics in a CPT-odd
  Lorentz-violating background}

\author{E. O. Silva\inst{1} \and F. M. Andrade\inst{2}}
\shortauthor{Silva and Andrade}

\institute{
  \inst{1}
  Departamento de F\'{i}sica,
  Universidade Federal do  Maranh\~{a}o -
  Campus Universit\'{a}rio do Bacanga,
  65085-580 S\~{a}o Lu\'{i}s-MA, Brazil\\
  \inst{2}
  Departamento de Matem\'{a}tica e Estat\'{i}stica,
  Universidade Estadual de Ponta Grossa -
  84030-900 Ponta Grossa-PR, Brazil
}


\abstract{
The Aharonov-Casher problem in the presence of a
Lorentz-violating background nonminimally coupled to a spinor
and a gauge field is examined.
Using an approach based on the self-adjoint extension method, an
expression for the bound state energies is obtained in terms of
the physics of the problem by determining the self-adjoint
extension parameter.
}

\date{\today}

\begin{document}

\maketitle

Since the construction of the standard model extension (SME),
proposed by Colladay and Kosteleck\'{y}
\cite{PRD.1997.55.6760,PRD.1998.58.116002,PRD.2004.69.105009}
(see also
\cite{PRD.1999.59.116008,PRL.1989.63.224,PRL.1991.66.1811})
quantum field theory systems have been studied in the
presence of Lorentz symmetry violation.
The SME includes Lorentz-violating (LV) terms in all the sectors
of the minimal standard model, becoming a suitable tool to
address LV effects in distinct physical systems. 
Several investigations have been developed in the context of
this theoretical framework in the latest years, involving field
theories
\cite{PRL.1999.82.3572,PRL.1999.83.2518,PRD.1999.60.127901,
PRD.2001.63.105015,PRD.2001.64.046013,JPA.2003.36.4937,
PRD.2006.73.65015,PRD.2009.79.123503,PD.2010.239.942,
PRD.2012.86.065011,PRD.2008.78.125013,PRD.2011.84.076006,
EPL.2011.96.61001,PRD.2012.85.085023,PRD.2012.85.105001,
EPL.2012.99.21003,PRD.2011.84.045008},
aspects on the gauge sector of the SME 
\cite{NPB.2003.657.214,NPB.2001.607.247,PRD.1995.51.5961,
PRD.1998.59.25002,PLB.1998.435.449,PRD.2003.67.125011,
PRD.2009.80.125040}, 
quantum electrodynamics 
\cite{EPJC.2008.56.571,PRD.2010.81.105015,PRD.2011.83.045018,
EPJC.2012.72.2070,JPG.2012.39.125001,JPG.2012.39.35002},
and astrophysics 
\cite{PRD.2002.66.081302,AR.2009.59.245,PRD.2011.83.127702}. 
These many contributions have elucidated the effects induced by
Lorentz violation and served to set up stringent upper bounds on
the LV coefficients \cite{RMP.2011.83.11}.

Another way to propose and investigate Lorentz violation is
considering new interaction terms.
In particular, in ref. \cite{EPJC.2005.41.421}, a
Lorentz-violating and CPT-odd nonminimal coupling between
fermions and the gauge field was firstly proposed in the
form
\begin{equation}
  D_{\mu }=\partial_{\mu }+ieA_{\mu }
  +i\frac{g}{2}\epsilon_{\mu \lambda\alpha \nu }
  V^{\lambda }F^{\alpha \nu },
  \label{eq:cov}
\end{equation}
in the context of the Dirac equation, 
$(i\gamma^{\mu}D_{\mu}-m)\Psi =0$.
In this case, the fermion spinor is $\Psi$, while
$\mathit{V}^{\mu }=(\mathit{V}_{0},\mathbf{V})$ is the
Carroll-Field-Jackiw four-vector, and $g$ is the constant that
measures the nonminimal coupling magnitude.
The analysis of the nonrelativistic limit of Eq. \eqref{eq:cov}
revealed that this nonminmal coupling generates a magnetic dipole moment
$(g\mathbf{V})$ even for uncharged particles
\cite{EPJC.2005.41.421}, yielding an Aharonov-Casher (AC) phase
for its wave function.
In these works, after assessing the nonrelativistic regime, one
has identified a generalized canonical momentum,
\begin{equation}
  \boldsymbol{\pi}=\mathbf{p}-e\mathbf{A}+g\mathit{V}_{0}
  \mathbf{B}-g\mathbf{V}\times \mathbf{E},
\label{eq:mto}
\end{equation}
which allows to introduce this nonminimal coupling in an
operational way, \textit{i.e.}, just redefining the vector potential and
the corresponding magnetic field as indicated below:
\begin{equation}
  \mathbf{A}\rightarrow
  \mathbf{A}+\frac{g}{e}
  \left(\mathbf{V}\times\mathbf{E}\right),  
\label{eq:a1}
\end{equation}
\begin{equation}
  \mathbf{B}=\boldsymbol{\nabla}\times\mathbf{A}
  \rightarrow \boldsymbol{\nabla}\times \mathbf{A}
  - \frac{g}{e}\boldsymbol{\nabla}
  \times \left(\mathbf{V}\times \mathbf{E}\right) .
  \label{eq:a2}
\end{equation}
This CPT-odd nonminimal coupling was further analyzed in various
contexts in relativistic quantum mechanics 
\cite{PRD.2006.74.065009,PRD.2012.86.045001,ADP.2011.523.910,
JMP.2011.52.063505,PRD.2011.83.125025,PLB.2006.639.675,
EPJC.2009.62.425,JPG.2012.39.055004,JPG.2012.39.105004}.

The aim of the present work is to study the effect of 
this CPT-odd LV nonminimal interaction on the AC dynamics.
Taking into account Eqs. \eqref{eq:a1} and \eqref{eq:a2} we
obtain the Schr\"{o}dinger-Pauli equation
\begin{equation}
\hat{H}\Psi =E\Psi,
  \label{eq:dedfm}
\end{equation}
where
\begin{align}
  \hat{H} = {} & \frac{1}{2M}
  \bigg\{
    \left[
    \mathbf{p}-
    e(\mathbf{A}+\frac{g}{e}\mathbf{V}\times \mathbf{E})
    \right]^{2}
    + e U\left( r\right)
  \nonumber \\
  {} &
    -e\,\boldsymbol{\sigma}\cdot
    \left[\boldsymbol{\nabla}\times\mathbf{A}
      -\frac{g}{e}\boldsymbol{\nabla}\times
      (\mathbf{V}\times\mathbf{E})
    \right]
  \bigg\},
  \label{eq:hdef}
\end{align}
is the Hamiltonian operator.
As is well-known, the potential $U(r)$, in cylindrical
coordinates, is given by
\begin{equation}
U(r) =-\phi\ln \left(\frac{r}{r_{0}}\right) .
\label{eq:ptln}
\end{equation}
It is very difficult to solve Eq. \eqref{eq:dedfm} 
by  including both effects taking into
account the logarithmic form of $U(r)$. 
This potential is relevant if we consider the full Hamiltonian,
which is compatible with a charged solenoid.
While the AC effect stems from the quantity
$\boldsymbol{\sigma}\cdot
[g\boldsymbol{\nabla}\times(\mathbf{V}\times\mathbf{E})]$,
one can affirm that the LV background does not
contribute to the Aharonov-Bohm (AB) effect.
To solve Eq. \eqref{eq:dedfm} considering both effects (AB
and AC) such potential has to be regarded.
Since we are interested only in the background effects,
we can examine only the sector generating the AC effect.
In this latter case, the field configuration is given by
\begin{equation}
  \mathbf{E}={\phi}\frac{\boldsymbol{\hat{r}}}{r},~~~
  \boldsymbol{\nabla}\cdot\mathbf{E}=\phi\frac{\delta (r)}{r},~~~
  \phi=\frac{\lambda}{2\pi \epsilon_{0}},~~~
  \mathbf{V}=V{\hat{\mathbf{z}}},
  \label{eq:acconf}
\end{equation}
where $\mathbf{E}$ is the electric field generated by an
infinite charge filament and $\lambda$ is the charge density
along the $z$-axis.
After this identification, the Hamiltonian \eqref{eq:hdef}
becomes
\begin{equation}
  \hat{H}=\frac{1}{2M}
  \left[
    \left(
      \frac{1}{i}\boldsymbol{\nabla}
      -\nu\frac{\hat{\mathbf{r}}}{r}
    \right)^{2}
    +\nu \sigma_{z} \frac{\delta(r)}{r}\right],
  \label{eq:hnrf}
\end{equation}
with
\begin{equation}
\nu = g V \phi,
\label{eq:deltac}
\end{equation}
the coupling constant of the $\delta(r)/r$ potential.
Here, we are only interested in the situation in
which $\hat{H}$ possesses bound states.

The Hamiltonian in Eq. \eqref{eq:hnrf} governs the quantum
dynamics of a spin-1/2 neutral particle with a radial electric
field, \textit{i.e.}, a spin-1/2 AC problem, with $g\mathbf{V}$ playing
the role of a nontrivial magnetic dipole moment.
Note the presence of a $\delta$ function singularity at the origin
in Eq. \eqref{eq:hnrf} which turns it more complicated to be
solved.
Such kind of point interaction potential can then be addressed by
the self-adjoint extension approach \cite{Book.2004.Albeverio},
used here for determining the bound states.

Making use of the underlying rotational symmetry expressed by the
fact that $[\hat{H},\hat{J}_{z}]=0$, where
$\hat{J}_{z}=-i\partial/\partial_{\phi}+\sigma_{z}/2$ is the total
angular momentum operator in the $z$-direction, we decompose the
Hilbert space $\mathfrak{H}=L^{2}(\mathbb{R}^{2})$
with respect to the angular momentum
$\mathfrak{H}=\mathfrak{H}_{r}\otimes\mathfrak{H}_{\varphi}$, where
$\mathfrak{H}_{r}=L^{2}(\mathbb{R}^{+},rdr)$ and
$\mathfrak{H}_{\varphi}=L^{2}(\mathcal{S}^{1},d\varphi)$, with
$\mathcal{S}^{1}$ denoting the unit sphere in $\mathbb{R}^{2}$.
So it is possible to express the eigenfunctions of the two
dimensional Hamiltonian in terms of the eigenfunctions of
$\hat{J}_{z}$:
\begin{equation}
  \Psi(r,\varphi)=
  \left(
    \begin{array}{c}
      \psi_{m}(r) e^{i(m_{j}-1/2)\varphi } \\
      \chi_{m}(r) e^{i(m_{j}+1/2)\varphi }
    \end{array}
  \right) ,
\label{eq:wavef}
\end{equation}
with $m_{j}=m+1/2=\pm 1/2,\pm 3/2,\ldots $, with $m\in
\mathbb{Z}$.
By inserting Eq. \eqref{eq:wavef} into Eq. \eqref{eq:dedfm} the
Schr\"{o}dinger-Pauli equation for $\psi_{m}(r)$ is found to be
\begin{equation}
H\psi_{m}(r)=E\psi_{m}(r),  \label{eq:eigen}
\end{equation}
where
\begin{equation}
H=H_{0}+\frac{\nu}{2M} \frac{\delta(r)}{r},
\label{eq:hfull}
\end{equation}
and
\begin{equation}
  H_{0}=-\frac{1}{2M}
  \left[
    \frac{d^{2}}{dr^{2}}+\frac{1}{r}\frac{d}{dr}
    -\frac{(m-\nu)^{2}}{r^{2}}
  \right].
  \label{eq:hzero}
\end{equation}

An operator $\mathcal{O}$, with domain
$\mathcal{D}(\mathcal{O})$, is said to be self-adjoint
if and only if
$\mathcal{D}(\mathcal{O}^{\dagger})=\mathcal{D}(\mathcal{O})$
and $\mathcal{O}^{\dagger}=\mathcal{O}$.
For smooth functions, $\xi \in C_{0}^{\infty}(\mathbb{R}^2)$ with
$\xi(0)=0$, we should have $H \xi = H_{0} \xi$.
Hence, it is reasonable to interpret the Hamiltonian
\eqref{eq:hfull} as a self-adjoint extension of
$H_{0}|_{C_{0}^{\infty}(\mathbb{R}^{2}\setminus \{0\})}$
\cite{crll.1987.380.87,JMP.1998.39.47,LMP.1998.43.43}.
It is a well-known fact that the symmetric radial operator
$H_{0}$ is essentially self-adjoint for $|m-\nu|\geq 1$
\cite{Book.1975.Reed.II}.
For those values of the $m$ fulfilling
$|m-\nu|<1$ it is not essentially self-ajoint, admitting an
one-parameter family of self-adjoint extensions, $H_{\theta,0}$,
where $\theta\in [0,2\pi)$ is the self-adjoint extension
parameter.
To characterize this family and determine the bound state
energy, we will follow a general approach proposed in
ref. \cite{PRD.2012.85.041701} (cf. also refs.
\cite{AP.2008.323.3150,AP.2010.325.2529,JMP.2012.53.122106,
PLB.2013.719.467}),
which is based on the boundary conditions 
that hold at the origin
\cite{CMP.1991.139.103}.
The boundary condition is a match of the logarithmic derivatives
of the zero-energy ($E=0$) solutions for 
Eq. \eqref{eq:eigen} and the solutions for the problem
defined by $H_{0}$ plus the self-adjoint extension.

Now, the goal is to find the bound states for the Hamiltonian
\eqref{eq:hfull}.
Then, we temporarily forget the $\delta$ function potential and
find the boundary conditions allowed for $H_{0}$.
However, the self-adjoint extension provides an infinity of
possible boundary conditions, and it can not give us the true
physics of the problem.
Nevertheless, once the physics at $r=0$ is known
\cite{AP.2008.323.3150,AP.2010.325.2529}, it is possible to
determine any arbitrary parameter coming from the self-adjoint
extension, and then we have a complete description of the
problem.
Since we have a singular point we must guarantee that the
Hamiltonian is self-adjoint in the region of motion.

One observes that even if the operator is Hermitian
$H_{0}^{\dagger}=H_{0}$, its domains could be different.
The self-adjoint extension approach consists, essentially, in
extending the domain $\mathcal{D}(H_{0})$ to match
$\mathcal{D}(H_{0}^{\dagger})$, therefore turning $H_{0}$
self-adjoint.
To do this, we must find the the deficiency subspaces,
$N_{\pm}$, with dimensions $n_{\pm}$, which are called
deficiency indices of $H_{0}$
\cite{Book.1975.Reed.II}.
A necessary and sufficient condition for $H_{0}$ being
essentially self-adjoint is that $n_{+}=n_{-}=0$.
On the other hand, if $n_{+}=n_{-}\geq 1$, then $H_{0}$ has an
infinite number of self-adjoint extensions parametrized by
unitary operators  $U:N_{+}\to N_{-}$.
In order to find the deficiency subspaces of $H_{0}$ in
$\mathfrak{H}_{r}$, we must solve the eigenvalue equation
\begin{equation}
  H_{0}^{\dagger}\psi_{\pm}
  =\pm i k_{0} \psi_{\pm},
  \label{eq:eigendefs}
\end{equation}
where $k_{0}\in \mathbb{R}$ is introduced for dimensional
reasons.
Since $H_{0}^{\dagger }=H_{0}$, the only square-integrable
functions which are solutions of Eq. \eqref{eq:eigendefs} are
the modified Bessel functions of second kind,
\begin{equation}
\psi_{\pm}=K_{|m-\nu|}(r\sqrt{\mp \varepsilon }),
\end{equation}
with $\varepsilon=2iM k_{0}$.
These functions are square integrable only in the range
$|m-\nu|<1$, for which $H_{0}$ is not self-adjoint.
The dimension of such deficiency subspace is
$(n_{+},n_{-})=(1,1)$.
According to the von Neumann-Krein theory, the domain of
$H_{\theta,0}$ is given by
\begin{equation}
\mathcal{D}(H_{\theta,0})=\mathcal{D}(H_{0}^{\dagger})=
\mathcal{D}(H_{0})\oplus N_{+}\oplus N_{-}.
\end{equation}
Thus, $\mathcal{D}(H_{\theta,0})$ in $\mathfrak{H}_{r}$ is given
by the set of functions \cite{Book.1975.Reed.II}
\begin{align}
\label{eq:domain}
\psi_{\theta}(r)={}& \psi_{m}(r)\nonumber \\
{} & +c\left[ K_{|m-\nu|}(r\sqrt{-\varepsilon})+
e^{i\theta}K_{|m-\nu|}(r\sqrt{\varepsilon}) \right] ,
\end{align}
where $\psi_{m}(r)$, with $\psi_{m}(0)=\dot{\psi}_{m}(0)=0$
($\dot{\psi}\equiv d\psi/dr$), is a regular wave function
and $\theta \in [0,2\pi)$ represents a choice for the boundary
condition.

Now, we are in position to determine a fitting value for
$\theta$.
To do so, we follow the approach of
ref. \cite{CMP.1991.139.103}.
First, one considers the zero-energy solutions $\psi_{0}$ and
$\psi_{\theta,0}$ for $H$ and $H_{0}$, respectively, \textit{i.e.},
\begin{equation}
  \left[
    \frac{d^{2}}{dr^{2}}+\frac{1}{r}\frac{d}{dr}-
    \frac{(m-\nu)^{2}}{r^{2}}-\nu \frac{\delta(r)}{r}
\right] \psi_{0}=0,
  \label{eq:statictrue}
\end{equation}
and
\begin{equation}
  \left[
    \frac{d^{2}}{dr^{2}}+\frac{1}{r}\frac{d}{dr}-
    \frac{(m-\nu)^{2}}{r^{2}}
  \right] \psi_{\theta,0}=0.
  \label{eq:thetastatic}
\end{equation}
The value of $\theta$ is determined by the boundary condition
\begin{equation}
  \lim_{a\to 0^{+}}
  \left(
    a\frac{\dot{\psi}_{0}}{\psi_{0}}\Big|_{r=a}-
    a\frac{\dot{\psi}_{\theta,0}}{\psi_{\theta,0}}\Big|_{r=a}
  \right)=0.
  \label{eq:logder}
\end{equation}
The first term of \eqref{eq:logder} is obtained by integrating
\eqref{eq:statictrue} from 0 to $a$.
The second term is calculated using the asymptotic
representation for the Bessel function $K_{|m-\nu|}$ for small
argument.
So, from \eqref{eq:logder} we arrive at
\begin{equation}
  \frac{\dot{\Upsilon}_{\theta}(a)}{\Upsilon_{\theta}(a)}=\nu,
  \label{eq:saepapprox}
\end{equation}
with
\begin{equation}
\Upsilon_{\theta}(r)=D(-\varepsilon)+e^{i\theta} D(+\varepsilon) ,
\end{equation}
and
\begin{equation}
  D(\pm \varepsilon)=
  \frac
  {\left(r\sqrt{\pm \varepsilon}\right)^{-|m-\nu|}}
  {2^{-|m-\nu|}\Gamma(1-|m-\nu|)}
  -\frac
  {\left(r\sqrt{\pm \varepsilon}\right)^{|m-\nu|}}
  {2^{|m-\nu|}\Gamma(1+|m-\nu|)}.
\end{equation}
Eq. \eqref{eq:saepapprox} gives us the parameter $\theta$ in
terms of the physics of the problem, \textit{i.e.}, the correct behavior
of the wave functions at the origin.

Next, we will find the bound states of the Hamiltonian $H_{0}$
and, by using \eqref{eq:saepapprox},
the spectrum of $H$ will be
determined without any arbitrary parameter.
Then, from $H_{0}\psi_{\theta}=E\psi_{\theta}$ we
achieve the modified Bessel equation
($\kappa^{2}=-2ME$)
\begin{equation}
  \left[
    \frac{d^{2}}{dr^{2}}+\frac{1}{r}\frac{d}{dr}-
    \frac{(m-\nu)^{2}}{r^{2}}-\kappa ^{2}
  \right]\psi_{\theta}(r)=0,
  \label{eq:eigenvalue}
\end{equation}
where $E<0$  (since we are looking for bound states).
The general solution for the above equation is
\begin{equation}
\psi_{\theta}(r)=K_{|m-\nu|}\left(r\sqrt{-2ME}\right).
\label{eq:sver}
\end{equation}
Since these solutions belong to $\mathcal{D}(H_{\theta,0})$,
they present the form \eqref{eq:domain} for a
$\theta$ selected from 
the physics of the problem (cf. Eq. \eqref{eq:saepapprox}).
So, we substitute \eqref{eq:sver} into \eqref{eq:domain} and
compute $a{\dot{\psi}_{\theta}}/{\psi_{\theta}}|_{r=a}$.
After a straightforward calculation, we have the relation
\begin{equation}
  \frac
  {|m-\nu|
    \left[
      a^{2|m-\nu|}(-ME)^{|m-\nu|}\Theta-1
    \right]
  }
  {a^{2|m-\nu|}(-ME)^{|m-\nu|}\Theta+1}
    =\nu,
\end{equation}
where $\Theta=\Gamma(-|m-\nu|)/(2^{|m-\nu|}\Gamma(|m-\nu|))$.
Solving the above equation for $E$, we find the sought
energy spectrum
\begin{equation}
  E=
  -\frac{2}{M a^{2}}
  \left[
    \left(
      \frac
      {\nu +|m-\nu|}
      {\nu -|m-\nu|}
    \right)
    \frac
    {\Gamma (1+|m-\nu|)}
    {\Gamma (1-|m-\nu|)}
  \right]^{{1}/{|m-\nu|}}.
\label{eq:energy_KS}
\end{equation}
In the above relation, to ensure that the energy is a real
number, we must have $|\nu| \geq |m-\nu|$, and due to $|m-\nu|<1$
it is sufficient to consider $|\nu|\geq 1$.
A necessary condition for a $\delta$ function generating an
attractive potential, able to support bound states, is
that the coupling constant must be negative.
Thus, the existence of bound states with real energies requires
\begin{equation}
\nu \leq -1.
\end{equation}
From the above equation and Eq \eqref{eq:deltac}  it
follows that $g V \lambda < 0$,
and there is a minimum value for this product.

In conclusion, we have analyzed the effects of a LV background
vector, nonminimally coupled to the gauge and fermion fields, on
the AC problem.
The self-adjoint extension approach was used to
determine the bound states of the 
particle in terms of the physics of the problem, in a very
consistent way and without any arbitrary parameter.

\acknowledgments
The authors would like to thank M. M. Ferreira Jr.
for critical reading the manuscript and helpful discussions.
E. O. Silva acknowledges researcher grants by CNPq-(Universal)
project No. 484959/2011-5.

\bibliographystyle{eplbib}

\begin{thebibliography}{10}
\expandafter\ifx\csname url\endcsname\relax\def\url#1{\texttt{#1}}\fi

\bibitem{PRD.1997.55.6760}
\Name{Colladay D. \and Kosteleck\'{y} V.~A.} \REVIEW{Phys. Rev.
  D}{55}{1997}{6760}.

\bibitem{PRD.1998.58.116002}
\Name{Colladay D. \and Kosteleck\'{y} V.~A.} \REVIEW{Phys. Rev.
  D}{58}{1998}{116002}.

\bibitem{PRD.2004.69.105009}
\Name{Kosteleck\'{y} V.~A.} \REVIEW{Phys. Rev. D}{69}{2004}{105009}.

\bibitem{PRD.1999.59.116008}
\Name{Coleman S. \and Glashow S.~L.} \REVIEW{Phys. Rev. D}{59}{1999}{116008}.

\bibitem{PRL.1989.63.224}
\Name{Kosteleck\'{y} V.~A. \and Samuel S.} \REVIEW{Phys. Rev.
  Lett.}{63}{1989}{224}.

\bibitem{PRL.1991.66.1811}
\Name{Kosteleck\'{y} V.~A. \and Samuel S.} \REVIEW{Phys. Rev.
  Lett.}{66}{1991}{1811}.

\bibitem{PRL.1999.82.3572}
\Name{Jackiw R. \and Kosteleck\'{y} V.~A.} \REVIEW{Phys. Rev.
  Lett.}{82}{1999}{3572}.

\bibitem{PRL.1999.83.2518}
\Name{P\'{e}rez-Victoria M.} \REVIEW{Phys. Rev. Lett.}{83}{1999}{2518}.

\bibitem{PRD.1999.60.127901}
\Name{Chung J.-M.} \REVIEW{Phys. Rev. D}{60}{1999}{127901}.

\bibitem{PRD.2001.63.105015}
\Name{Chung J.-M. \and Chung B.~K.} \REVIEW{Phys. Rev. D}{63}{2001}{105015}.

\bibitem{PRD.2001.64.046013}
\Name{Scarpelli A. P.~B., Sampaio M., Nemes M.~C. \and Hiller B.} \REVIEW{Phys.
  Rev. D}{64}{2001}{046013}.

\bibitem{JPA.2003.36.4937}
\Name{Bazeia D., Mariz T., Nascimento J.~R., Passos E. \and Ribeiro R.~F.}
  \REVIEW{J. Phys. A}{36}{2003}{4937}.

\bibitem{PRD.2006.73.65015}
\Name{Barreto M.~N., Bazeia D. \and Menezes R.} \REVIEW{Phys. Rev.
  D}{73}{2006}{065015}.

\bibitem{PRD.2009.79.123503}
\Name{Avelino P.~P., Bazeia D., Losano L., Menezes R. \and Rodrigues J.~J.}
  \REVIEW{Phys. Rev. D}{79}{2009}{123503}.

\bibitem{PD.2010.239.942}
\Name{Bazeia D., Ferreira~Jr. M.~M., Gomes A.~R. \and Menezes R.} \REVIEW{Phys.
  D}{239}{2010}{942}.

\bibitem{PRD.2012.86.065011}
\Name{Miller C., Casana R., Ferreira~Jr. M.~M. \and da~Hora E.} \REVIEW{Phys.
  Rev. D}{86}{2012}{065011}.

\bibitem{PRD.2008.78.125013}
\Name{Casana R., Ferreira~Jr. M.~M. \and Rodrigues J.~S.} \REVIEW{Phys. Rev.
  D}{78}{2008}{125013}.

\bibitem{PRD.2011.84.076006}
\Name{Altschul B.} \REVIEW{Phys. Rev. D}{84}{2011}{076006}.

\bibitem{EPL.2011.96.61001}
\Name{Ganguly O., Gangopadhyay D. \and Majumdar P.} \REVIEW{Europhys.
  Lett.}{96}{2011}{61001}.

\bibitem{PRD.2012.85.085023}
\Name{Cambiaso M., Lehnert R. \and Potting R.} \REVIEW{Phys. Rev.
  D}{85}{2012}{085023}.

\bibitem{PRD.2012.85.105001}
\Name{Pospelov M. \and Shang Y.} \REVIEW{Phys. Rev. D}{85}{2012}{105001}.

\bibitem{EPL.2012.99.21003}
\Name{Leite J. \and Mariz T.} \REVIEW{Europhys. Lett.}{99}{2012}{21003}.

\bibitem{PRD.2011.84.045008}
\Name{Casana R., Carvalho E.~S. \and Ferreira~Jr. M.~M.} \REVIEW{Phys. Rev.
  D}{84}{2011}{045008}.

\bibitem{NPB.2003.657.214}
\Name{Adam C. \and Klinkhamer F.} \REVIEW{Nucl. Phys. B}{657}{2003}{214}.

\bibitem{NPB.2001.607.247}
\Name{Adam C. \and Klinkhamer F.} \REVIEW{Nucl. Phys. B}{607}{2001}{247}.

\bibitem{PRD.1995.51.5961}
\Name{Andrianov A.~A. \and Soldati R.} \REVIEW{Phys. Rev. D}{51}{1995}{5961}.

\bibitem{PRD.1998.59.25002}
\Name{Andrianov A.~A., Soldati R. \and Sorbo L.} \REVIEW{Phys. Rev.
  D}{59}{1998}{025002}.

\bibitem{PLB.1998.435.449}
\Name{Andrianov A. \and Soldati R.} \REVIEW{Phys. Lett. B}{435}{1998}{449}.

\bibitem{PRD.2003.67.125011}
\Name{Belich H., Ferreira~Jr. M.~M., Helay\"el-Neto J.~A. \and Orlando M.
  T.~D.} \REVIEW{Phys. Rev. D}{67}{2003}{125011}.

\bibitem{PRD.2009.80.125040}
\Name{Casana R., Ferreira~Jr. M.~M., Gomes A.~R. \and Pinheiro P. R.~D.}
  \REVIEW{Phys. Rev. D}{80}{2009}{125040}.

\bibitem{EPJC.2008.56.571}
\Name{Scarpelli A., Sampaio M., Nemes M. \and Hiller B.} \REVIEW{Eur. Phys. J.
  C}{56}{2008}{571}.

\bibitem{PRD.2010.81.105015}
\Name{Casana R., Ferreira~Jr. M.~M. \and Silva M. R.~O.} \REVIEW{Phys. Rev.
  D}{81}{2010}{105015}.

\bibitem{PRD.2011.83.045018}
\Name{Mariz T.} \REVIEW{Phys. Rev. D}{83}{2011}{045018}.

\bibitem{EPJC.2012.72.2070}
\Name{Casana R., Ferreira~Jr. M.~M. \and Moreira R. P.~M.} \REVIEW{Eur. Phys.
  J. C}{72}{2012}{2070}.

\bibitem{JPG.2012.39.125001}
\Name{Scarpelli A. P.~B.} \REVIEW{J. Phys. G}{39}{2012}{125001}.

\bibitem{JPG.2012.39.35002}
\Name{Gazzola G., Fargnoli H.~G., Scarpelli A. P.~B., Sampaio M. \and Nemes
  M.~C.} \REVIEW{J. Phys. G}{39}{2012}{035002}.

\bibitem{PRD.2002.66.081302}
\Name{Jacobson T., Liberati S. \and Mattingly D.} \REVIEW{Phys. Rev.
  D}{66}{2002}{081302}.

\bibitem{AR.2009.59.245}
\Name{Liberati S. \and Maccione L.} \REVIEW{Ann. Rev. Nuclear Particle
  Science}{59}{2009}{245}.

\bibitem{PRD.2011.83.127702}
\Name{S. L. \and Ma B.} \REVIEW{Phys. Rev. D}{83}{2011}{127702}.

\bibitem{RMP.2011.83.11}
\Name{Kosteleck\'{y} V.~A. \and Russell N.} \REVIEW{Rev. Mod.
  Phys.}{83}{2011}{11}.

\bibitem{EPJC.2005.41.421}
\Name{Belich H., Costa-Soares T., Ferreira~Jr. M.~M. \and Helay\"{e}l-Neto
  J.~A.} \REVIEW{Eur. Phys. J. C}{41}{2005}{421}.

\bibitem{PRD.2006.74.065009}
\Name{Belich H., Costa-Soares T., Ferreira~Jr. M.~M., Helay\"{e}l-Neto J.~A.
  \and Moucherek F. M.~O.} \REVIEW{Phys. Rev. D}{74}{2006}{065009}.

\bibitem{PRD.2012.86.045001}
\Name{Grushin A.~G.} \REVIEW{Phys. Rev. D}{86}{2012}{045001}.

\bibitem{ADP.2011.523.910}
\Name{Bakke K., Belich H. \and Silva E.~O.} \REVIEW{Ann. Phys.
  (Berlin)}{523}{2011}{910}.

\bibitem{JMP.2011.52.063505}
\Name{Bakke K., Belich H. \and Silva E.~O.} \REVIEW{J. Math.
  Phys.}{52}{2011}{063505}.

\bibitem{PRD.2011.83.125025}
\Name{Belich H., Silva E.~O., Ferreira~Jr. M.~M. \and Orlando M. T.~D.}
  \REVIEW{Phys. Rev. D}{83}{2011}{125025}.

\bibitem{PLB.2006.639.675}
\Name{Belich H., Costa-Soares T., Ferreira~Jr. M.~M., Helay\"{e}l-Neto J. \and
  Orlando M.} \REVIEW{Phys. Lett. B}{639}{2006}{675}.

\bibitem{EPJC.2009.62.425}
\Name{Belich H., Colatto L., Costa-Soares T., Helay\"{e}l-Neto J. \and Orlando
  M.} \REVIEW{Eur. Phys. J. C}{62}{2009}{425}.

\bibitem{JPG.2012.39.055004}
\Name{Bakke K., Silva E.~O. \and Belich H.} \REVIEW{J. Phys. G: Nucl. Part.
  Phys.}{39}{2012}{055004}.

\bibitem{JPG.2012.39.105004}
\Name{Ribeiro L.~R., Furtado C. \and Passos E.} \REVIEW{J. Phys.
  G}{39}{2012}{105004}.

\bibitem{Book.2004.Albeverio}
\Name{Albeverio S., Gesztesy F., Hoegh-Krohn R. \and Holden H.} \Book{Solvable
  Models in Quantum Mechanics} 2nd Edition (AMS Chelsea Publishing, Providence,
  RI) 2004.

\bibitem{crll.1987.380.87}
\Name{Gesztesy F., Albeverio S., Hoegh-Krohn R. \and Holden H.} \REVIEW{J.
  Reine Angew. Math.}{380}{1987}{87}.

\bibitem{JMP.1998.39.47}
\Name{Dabrowski L. \and \v{S}t'ov\'{i}\v{c}ek P.} \REVIEW{J. Math.
  Phys.}{39}{1998}{47}.

\bibitem{LMP.1998.43.43}
\Name{Adami R. \and Teta A.} \REVIEW{Lett. Math. Phys.}{43}{1998}{43}.

\bibitem{Book.1975.Reed.II}
\Name{Reed M. \and Simon B.} \Book{Methods of Modern Mathematical Physics. II.
  Fourier Analysis, Self-Adjointness.} (Academic Press, New York - London)
  1975.

\bibitem{PRD.2012.85.041701}
\Name{Andrade F.~M., Silva E.~O. \and Pereira M.} \REVIEW{Phys. Rev.
  D}{85}{2012}{041701(R)}.

\bibitem{AP.2008.323.3150}
\Name{Filgueiras C. \and Moraes F.} \REVIEW{Ann. Phys. (NY)}{323}{2008}{3150}.

\bibitem{AP.2010.325.2529}
\Name{Filgueiras C., Silva E.~O., Oliveira W. \and Moraes F.} \REVIEW{Ann.
  Phys. (NY)}{325}{2010}{2529}.

\bibitem{JMP.2012.53.122106}
\Name{Filgueiras C., Silva E.~O. \and Andrade F.~M.} \REVIEW{J. Math.
  Phys.}{53}{2012}{122106}.

\bibitem{PLB.2013.719.467}
\Name{Andrade F.~M. \and Silva E.~O.} \REVIEW{Phys. Lett. B}{719}{2013}{467}.

\bibitem{CMP.1991.139.103}
\Name{Kay B.~S. \and Studer U.~M.} \REVIEW{Commun. Math.
  Phys.}{139}{1991}{103}.

\end{thebibliography}

\end{document}